\documentclass[aps,prl,superscriptaddress,twocolumn,10pt,nofootinbib,preprintnumbers]{revtex4}

\usepackage{bm}
\usepackage{epsfig}
\usepackage{graphics}
\usepackage{amsmath}
\usepackage{xcolor}

\begin{document}

\title{Crossover of dynamical instability and chaos in the supercritical state}
\author{\firstname{C.}~\surname{Cockrell}} \email{c.j.cockrell@qmul.ac.uk}
\affiliation{School of Physics and Astronomy, Queen Mary University of London, Mile End Road, London, E1 4NS, UK}

\begin{abstract}
We calculate the maximal Lyapunov exponent for a bulk system of 256 Lennard-Jones particles in constant energy molecular dynamics simulations deep into the supercritical state. We find that the maximal Lyapunov exponent undergoes a crossover, and that this crossover coincides with the dynamical crossover at the Frenkel line from liquid physics. We explain this crossover in terms of two different contributions to dynamical instability: diffusion in the liquid-like state below the Frenkel line, and collisions in the gas-like state above. These results provide insight into the phase space dynamics far from the melting line and densities where rare-gas approximation are inapplicable.
\end{abstract}

\maketitle

\section{Introduction}

\begin{figure}
\begin{center}
{\scalebox{0.41}{\includegraphics{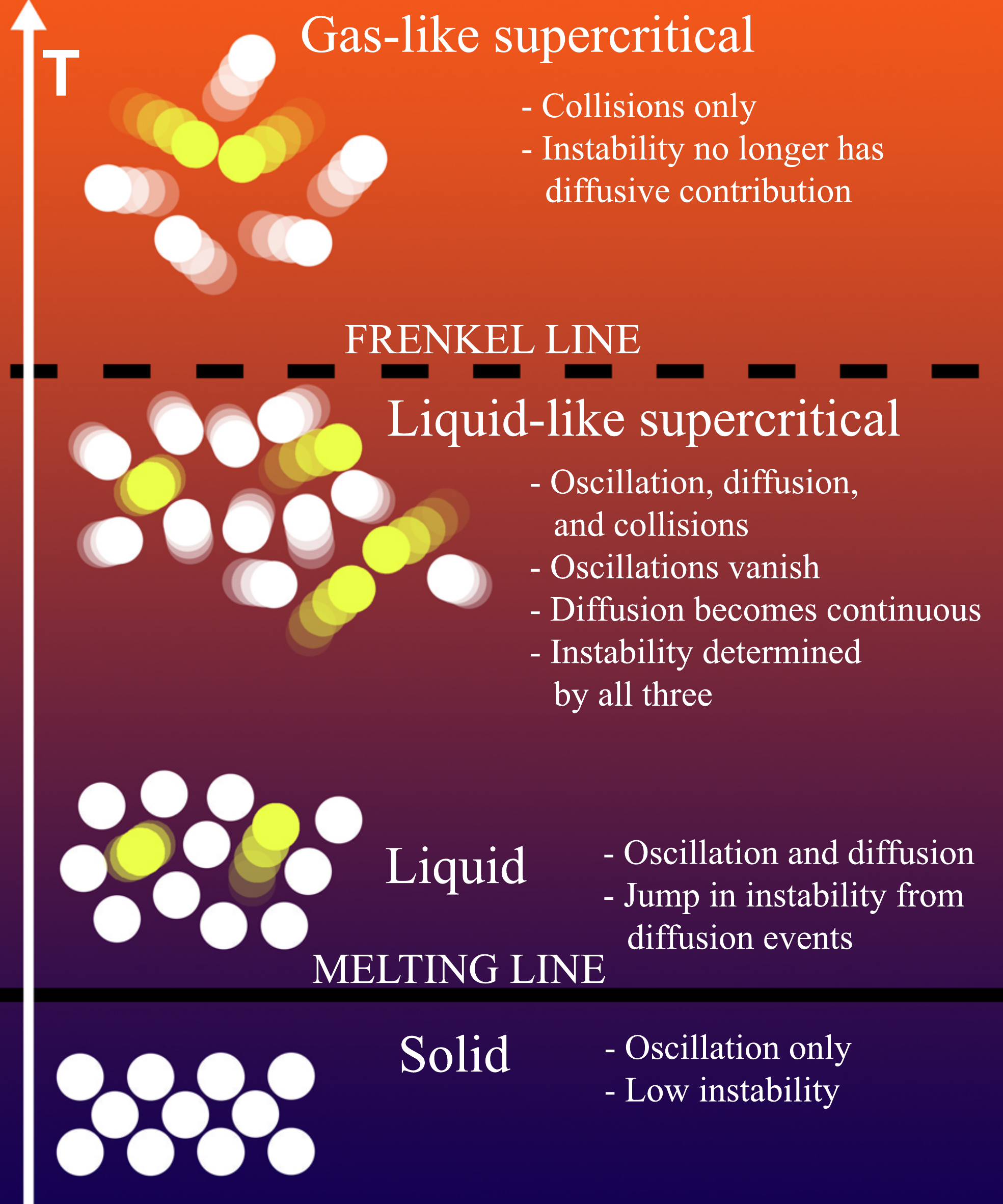}}}
\end{center}
\caption{Summary of our main results: Evolution of dynamical instability in condensed matter, from solids at low temperature to gas-like supercritical fluids at high temperature. The figure shows the dynamical regimes in each state of matter (oscillation, diffusion, collisions) and their relationship to the dynamical instability.}
\label{fig:diagram}
\end{figure}

The process of equilibration remains one of the major unanswered questions in nonequilibrium statistical mechanics \cite{Zwanzig2001}. This is despite the phenomenal success and ubiquity of equilibrium statistical mechanical methods in all fields of physics. The power of the ergodic hypothesis introduced by Boltzmann and Gibbs lies in the equality of empirical time averages and theoretical ensemble averages over initial conditions. An ergodic system requires that any given trajectory must, in the long term, visit the available phase space in a manner statistically independent of its initial state. This already suggests dynamical instability. The approach to equilibrium, however, requires an initial probability density to both relax into a time-independent form and to spread out over the available phase space. The possibility of such a change occurring irreversibly has been debated since Boltzmann's times \cite{Cercignani2006}, but since the middle of the last century the idea of coarse-grained degrees of freedom (be they environmental or ``fast"/microscopic) has received repeated attention \cite{Kirkwood1946,Green1952,Zwanzig1961,Mori1965,Evans1993,Evans1994,Evans2002,Sekimoto1998,Seifert2012}. The measure-preserving (Liouvillean) dynamics of Hamiltonian systems prohibits the diffusive smoothing of the probability density function into its equilibrium form starting from an arbitrary initial state \cite{Zwanzig2001}. However, in certain dynamical systems, any given small region of phase space can evolve under the dynamics to \textit{span} a far larger hypervolume while preserving its measure. This behaviour is expected of equilibrating probability densities and can be combined with some degree of coarse-graining to qualitatively produce the diffusive smoothing of a relaxing probability density. Such systems are said to have the \textit{mixing} property. 

Define a dynamical system $\mathcal{D} =  (\Gamma, T, \mu, \mathcal{M})$ with phase coordinates $\Gamma$, time-evolution operator $T(t)$, $\mu$ a measure (we might write $\rho$ if it's a probability density) and $\mathcal{M}$ the phase space. Then $\mathcal{D}$ has the mixing property if, for any $A, B \in \mathcal{M}$ \cite{Dorfman1999}

\begin{equation}
    \label{eq:mixingdef}
    \lim\limits_{t\to\infty} \frac{\mu(A_t \cup B)}{\mu(A_0)} = \frac{\mu(B)}{\mu(\mathcal{M})},
\end{equation}
\
where $A_0$ is the set $A$ at a starting time $t=0$ and $A_t = T(t) A_0$ is that set evolved forward by time $t$. In the case that the measure is a probability density $\mu(\mathcal{M})$ will be normalised to unity, yielding the perhaps more familiar expression of the definition. Intuitively, what this definition means is that any set will evolve after enough time to uniformly explore the available phase space such that the proportional ``overlap" between that set and any region of phase space is equal to the proportional ``overlap" between that region and the total phase space. This fulfills the requirement that all initial probability densities must eventually reach the same equilibrium state. This argument can be made quantitative - phase (and time) averaged properties in a mixing system approach well-defined constants as $t \to \infty$ \cite{Dorfman1999}.

The mixing property has its origins in dynamical instability \cite{Krylov1980}, of which two related measures are the Lyapunov spectrum and the Kolmogorov-Sinae entropy \cite{Dorfman1999}. Lyapunov spectra measure the rate of divergence of neighbouring trajectories in phase space. Consider a point $\Gamma(0) \in \mathcal{M}$ at time $t=0$ and its perturbation in the phase space direction $i$, $\Gamma(0) + \delta \Gamma_i(0)$. If the dynamics are unstable, this perturbation will rapidly grow and erase correlations between the two trajectories. The state after time $t=\tau$ can be written $\Gamma(\tau) + \delta \Gamma(\tau)$, where $\Gamma(\tau)$ is the time-evolved unperturbed trajectory, and the time-evolved perturbation $\delta \Gamma(\tau)$ will, in general, spread into all phase space dimensions. The Lyapunov exponents, $\lambda_i$ are defined \cite{Gaspard1998}
\
\begin{equation}
    \label{eq:lyapunovexpdef}
    \lambda_i = \lim\limits_{t\to\infty} \frac{1}{t} \log \left( \frac{\mid \delta \Gamma_i(t) \mid}{\mid \delta \Gamma_i(0) \mid} \right).
\end{equation}
\
The Lyapunov spectrum therefore defines the directions in which the phase space contracts and expands under time evolution. The sum of the Lyapunov exponents is related to the phase space contraction rate, and thus vanished for Liouvillean flows \cite{Gaspard1998}. Exponential divergence means the presence of just one positive Lyapunov exponent signifies dynamical instability, and the perturbation size $\frac{\mid \delta \Gamma(t) \mid}{\mid \delta \Gamma(0) \mid}$ will be dominated by the largest positive Lyapunov exponent, $\Lambda$. We may therefore write
\
\begin{equation}
    \label{eq:maxlyapunovexpdef}
    \Lambda = \lim\limits_{t\to\infty} \frac{1}{t} \log \left( \frac{\mid \delta \Gamma(t) \mid}{\mid \delta \Gamma(0) \mid} \right),
\end{equation}
\
assuming that we can ignore the contrived case where the initial perturbation is perpendicular to the fastest expanding phase space direction.

The Kolmogorov-Sinai (KS) entropy, which measures the rate at which information is lost to coarse-graining \cite{Gaspard1998,Zaslavsky1985}, is the sum of all positive Lyapunov exponents \cite{Shimada1979}. The KS entropy can be interpreted as speed at which an initial cell of phase space spreads across the entirety of the available phase space (while its measure remains constant), and therefore is related to the rate at which equilibrium is approached. The Lyapunov spectrum will, in general, be a function of the phase space $\mathcal{M}$, however for ergodic systems it can immediately be seen from equation (\ref{eq:lyapunovexpdef}) that the spectrum will be a constant for a given dynamical system.

The chaotic dynamics of systems has received ample attention from the perspective of irreversible physics beyond theoretical considerations. Lyapunov exponents and KS entropy have been used to calculate nonequilibrium transport properties \cite{Posch1988, Dorfman1995, Gaspard1990,Evans1990a}. Additionally, the ability of digital computers to faithfully represent the dynamics of chaotic systems is an increasing important question (for a particularly striking example of a digital computer's failure, see the recent work \cite{Boghosian2019}), and the Lyapunov spectrum has been proposed as a natural measure of the deviation of the calculated trajectory from the ``true" one \cite{Norman2013}. The Lyapunov spectrum is therefore of significant theoretical importance in several different areas.

\begin{figure}
\begin{center}
{\scalebox{0.41}{\includegraphics{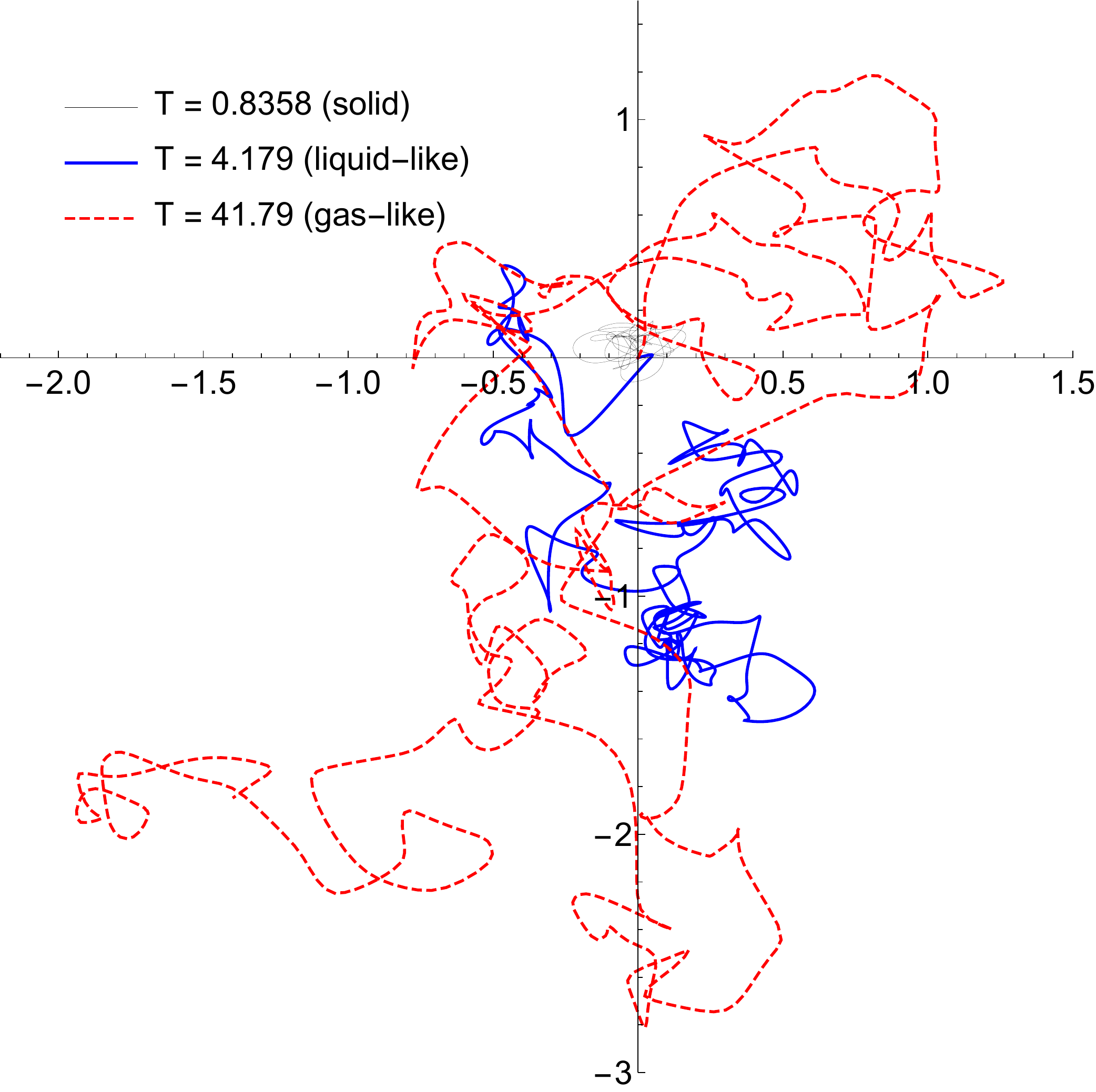}}}
\end{center}
\caption{Projected particle trajectories on the x-y plane (in reduced length units, see Tab. \ref{tab:ljparams}) at a reduced concentration of 1.056 in the solid phase, and the fluid phase below and above the Frenkel line.}
\label{fig:traj}
\end{figure}

In this work we study the maximal Lyaponuov exponent (MLE) $\Lambda$ of atomic Lennard-Jones (LJ) systems in  the solid phase and liquid phase above and below the Frenkel line (FL) using molecular dynamics (MD) simulations. The FL separates two different dynamic regimes in the fluid phase: at temperatures below the line, atomic motion combines oscillation around quasi-equilibrium positions with diffusive jumps (``liquid-like"; at temperatures above the line, atomic motion is purely diffusive (``gas-like") \cite{Brazhkin2012,Brazhkin2013,Trachenko2016}. See Fig. \ref{fig:traj} for representative trajectories in these different states from MD simulations. In the solid, atoms oscillate within a roughly fixed local environment. In the ``liquid-like" state below the FL, atoms spend some time oscillating, and some time diffusing between local environments. In the ``gas-like" state above the FL, atoms are continuously diffusive without interruption. Below the FL, therefore, the fluid supports a local rigid structure on small timescales. This gives a practical criterion to calculate the FL based on the disappearance of the minima of the velocity autocorrelation function (VAF). The FL represents not only a crossover in dynamics, but also in thermodynamics and structure \cite{Wang2017,Wang2019,Prescher2017,Smith2017,Proctor2019,Cockrell2020,Cockrell2020a}. The nature of the crossover at the FL is not yet well understood, and whether or not it's accompanied by a phase transition is still an open question. One of the seminal examples used in chaos theory is the Lorentz gas \cite{Bunimovich1980}, which models an ideal gas in the dilute limit, and whose Lyapunov exponent is well-known \cite{Gaspard1998}. On the other hand, the Lyapunov spectrum and KS entropy of condensed phases have also been well studied in the condensed phase using MD simulations \cite{Posch1989,Nayak1995,Mehra1997,Dellago1997,Kwon1997}. Furthermore, the behaviour of Lyapunov spectra across phase transitions has been documented \cite{Nayak1995, Mehra1997, Dellago1997,Kwon1997,Butera1987,Nayak1998,Barre2001}, exhibiting discontinuities in the MLE itself or its first derivative with respect to an order parameter. Phases are an ultimately macroscopic notion, but particle dynamics and phase space properties (Lyapunov spectra and KS entropy) can both provide a quantitative \textit{microscopic} description of the phases and the transitions between them, motivating our line of inquiry.

\section{Methods}

We consider a bulk system of 256 atoms with periodic boundary conditions interacting under the LJ potential:
\begin{equation}
    \label{eq:ljpotential}
    V_{ij} = 4 \epsilon \left( \left(\frac{\sigma}{r_{ij}}\right)^{12} - \left(\frac{\sigma}{r_{ij}}\right)^{6} \right),
\end{equation}
where $V_{ij}$ is the pair potential energy between atoms $i$ and $j$, $\epsilon$ is the well depth, $\sigma$ is the atomic radius, and $r_{ij}$ is the radial distance between them. For the purpose of continuity, we have selected the LJ potential characterising argon (see Tab. \ref{tab:ljparams}), whose behaviour across the FL has been well-characterised \cite{Wang2017, Yang2017, Wang2019}. We use the DL\_POLY MD simulation package \cite{Todorov2006}, which in the NVE ensemble uses the Velocity-Verlet \cite{Allen1991} integration algorithm. In simulations, we use time units of picoseconds (ps) and integration timesteps between $10^{-3}$ and $10^{-5}$ps, with not any of our results displaying sensitivity on the choice of timestep within this range. For analysis and discussion, we use the reduced time $t^{*} = t/\tau$ ($\tau = \sqrt{m \sigma^2 / \epsilon} = 2.163 \rm{ps}$) instead. Total energy is conserved to within 0.01\% for all production runs.

Our initial configuration consisted of 256 argon atoms in a cubic cell arranged in an FCC crystal structure \cite{Barrett1964} with lattice constants of 6.428, 6.049, 5.747, and 5.285 \rm{\AA}, corresponding to reduced concentrations ($n^* = n/\sigma^3$) concentrations of 0.5917, 0.7101, 0.8284, and 1.065 respectively. Mass and number densities in standard units are given in Tab. \ref{tab:tdparams}. Systems were then heated in the NVT ensemble using the Langevin thermostat \cite{Allen1991} with a relaxation time of 1.0 ps for $2\times10^5$ MD timesteps. The temperature ranged from 20 K in the crystalline state to 5000 K in the deep supercritical state, passing the melting line and Frenkel line. Temperature $T$ is defined from equipartiation as
\begin{equation}
    \label{eq:equipartition}
    T = \frac{2 \overline{E_{\rm{kin}}}}{(3N - 6) k_{\rm{B}}},
\end{equation}
where $\overline{E_{\rm{kin}}}$ is the kinetic energy averaged over the trajectory, $N$ is the number of atoms, and $k_{\rm{B}}$ is Boltzmann's constant. Near the melting line, we increased the density of temperature points to capture the sharp transition there. Relevant physical parameters in DL\_POLY units and reduced LJ units are listed in Tab. \ref{tab:tdparams}. The configurations generated in the equilibration stage were used as initial conditions for data collection in the NVE ensemble for $2\times10^5$ MD timesteps. From this stage statistical data such as total energy, diffusion coefficients, and VAFs were collected.

Following data collection were the production runs where the MLE were calculated. We calculated the MLE using the tangent space method \cite{Benettin1976}: At the beginning of the production run, the phase space was perturbed in such a way that every phase space coordinate is changed, but the total energy remains fixed. The system is evolved for a time $ \Delta t^* =$ 0.25 ( $\approx$ 0.5 ps) before the MLE is calculated using equation \ref{eq:maxlyapunovexpdef}. The evolved perturbation is then projected along itself such that its magnitude equals that of the initial perturbation $| \delta \Gamma(0)|$ and the process is repeated up to 100 times. The MLE we calculated this way is insensitive to our choices of initial perturbation size $| \delta \Gamma(0)|$ and the evolution time $\Delta t^*$ within reasonable ranges. The calculated values are then averaged to give the mean MLE \textit{over a given trajectory}, $\overline{\Lambda}$. We calculate identical results for $\overline{\Lambda}$ (up to statistical fluctuations) under different initial conditions. This fact, combined with states neighbouring in total energy having neighbouring values of $\overline{\Lambda}$ means that the time averaged MLE, $\overline{\Lambda}$, and the phase average MLE, $\langle \Lambda \rangle$, are the same quantity. From here on out we shall not distinguish these two quantities and use the term MLE and symbol $\Lambda$ to refer to  them.

\begin{table}
\begin{center}
\begin{tabular}{ |c|c|c| } 
\hline
Parameter & Value \\
 \hline
    mass (amu)  & 39.95 \\    
    $\epsilon$ (eV)  & 0.01032 \\   
    $\epsilon$ (K)  & 119.65 \\   
    $ \sigma$ (\AA) & 3.4 \\   
 \hline
\end{tabular}
\caption{Potential parameters used in the molecular dynamics simulations.}
\label{tab:ljparams}
\end{center}
\end{table}

\section{Results and Discussion}

\begin{figure}
\begin{center}
{\scalebox{0.41}{\includegraphics{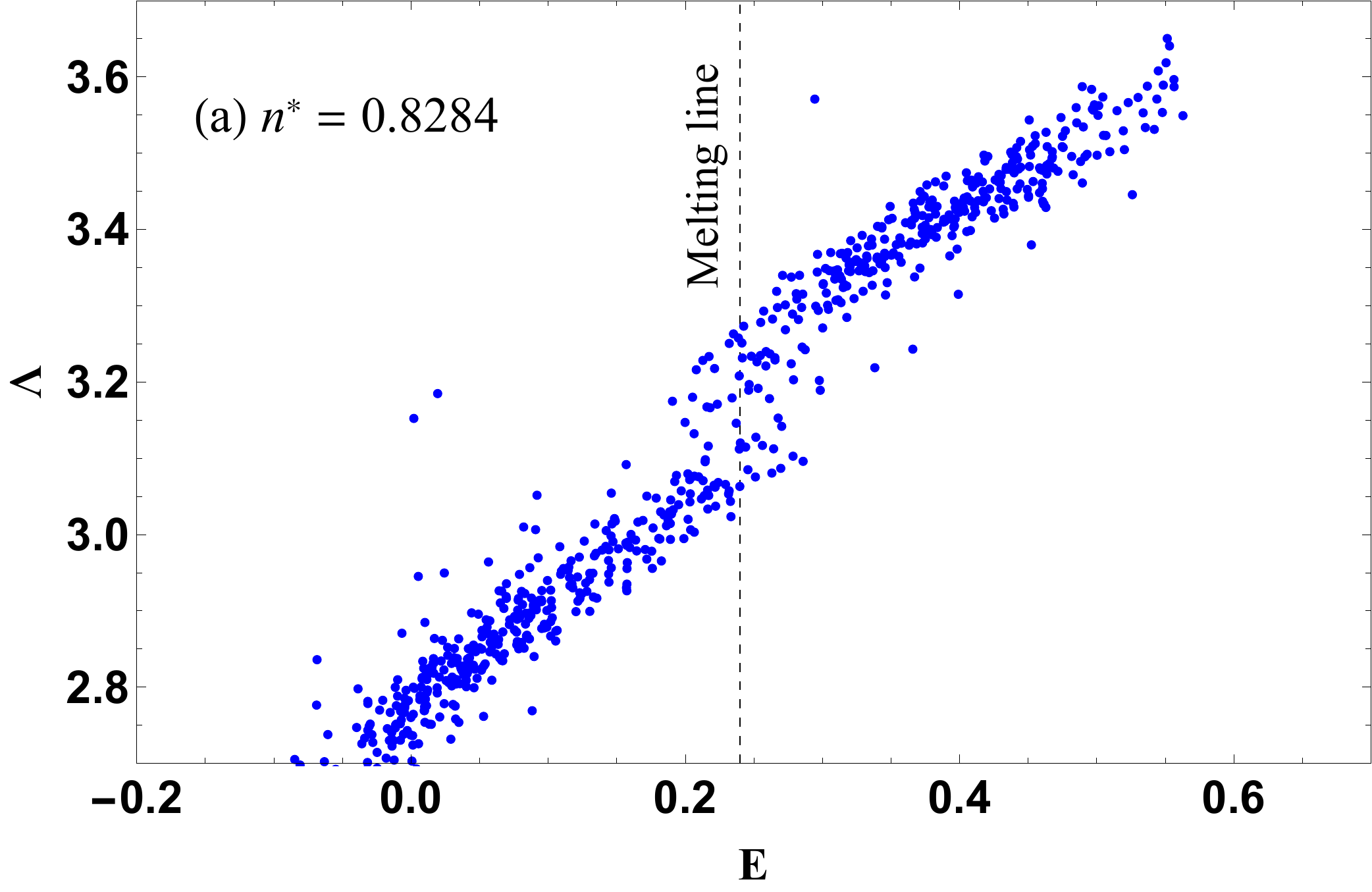}}}
{\scalebox{0.41}{\includegraphics{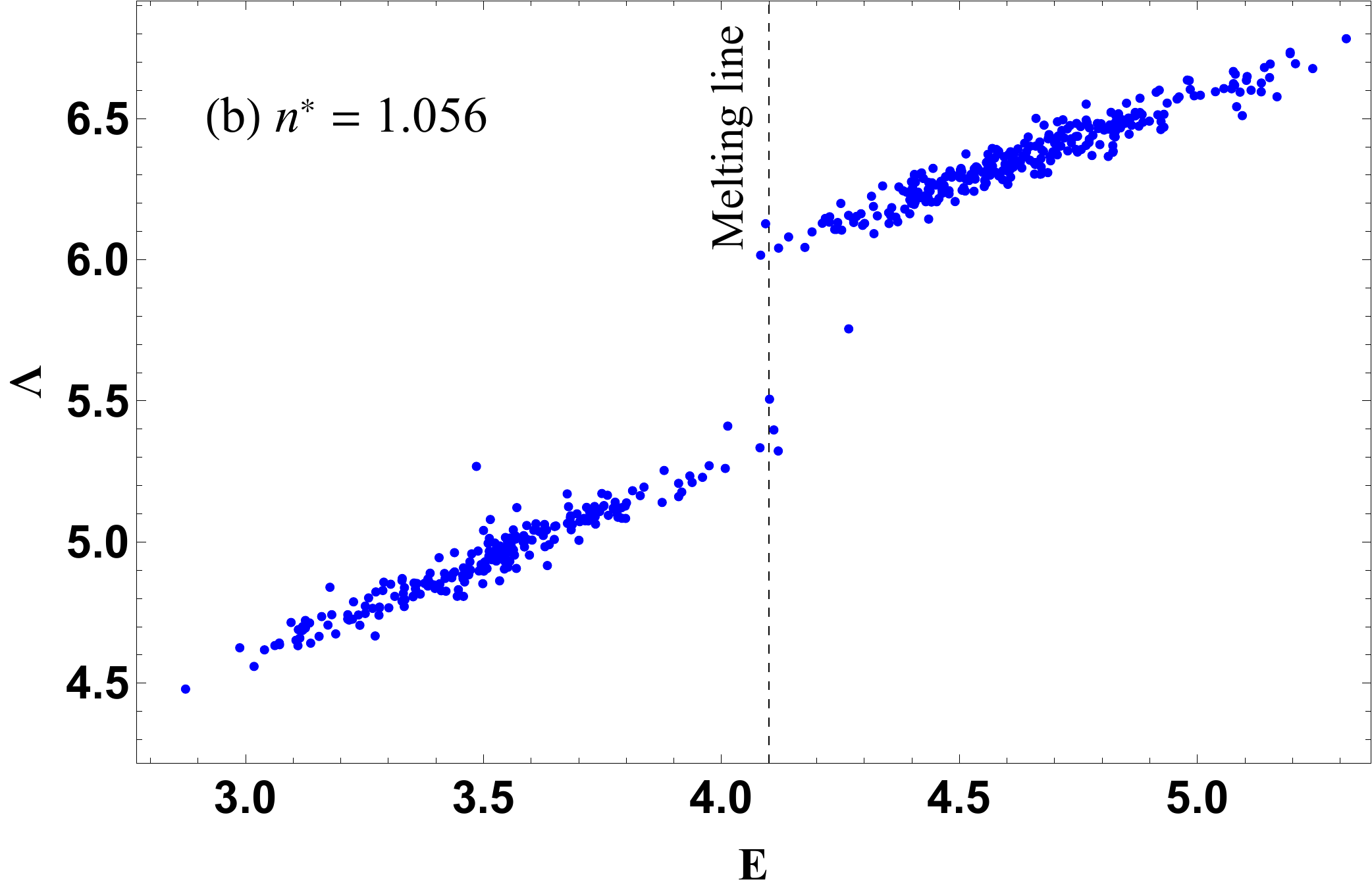}}}
\end{center}
\caption{Maximal Lyapunov exponent $\Lambda$ across the melting line at reduced concentrations of (a) 0.8284 and (b) 1.065.}
\label{fig:ptplot}
\end{figure}

We first discuss the transition at the melting line. Reduced energies as a function of reduced temperature are plotted in Fig. \ref{fig:ptplot}. The well-known discontinuity of energy across the melting line allows us to discern its location when we plot the MLE versus reduced energy in Fig. \ref{fig:ptplot}. We see the previously documented \cite{Nayak1995, Mehra1997, Dellago1997,Kwon1997,Nayak1998} discontinuity in the MLE in the transition from the solid to liquid phase. Dynamically speaking, the distinction between these phases is that liquids combine oscillation with diffusive jumps (in this sense the liquid is called a ``dynamically mixed state" \cite{Trachenko2016}). This was used by Nayak \textit{et. al.} to describe the discontinuity of the MLE in terms of the sudden expansion in the available phase space. This is a point we shall return to when we discuss the FL.

\begin{figure}
\begin{center}
{\scalebox{0.41}{\includegraphics{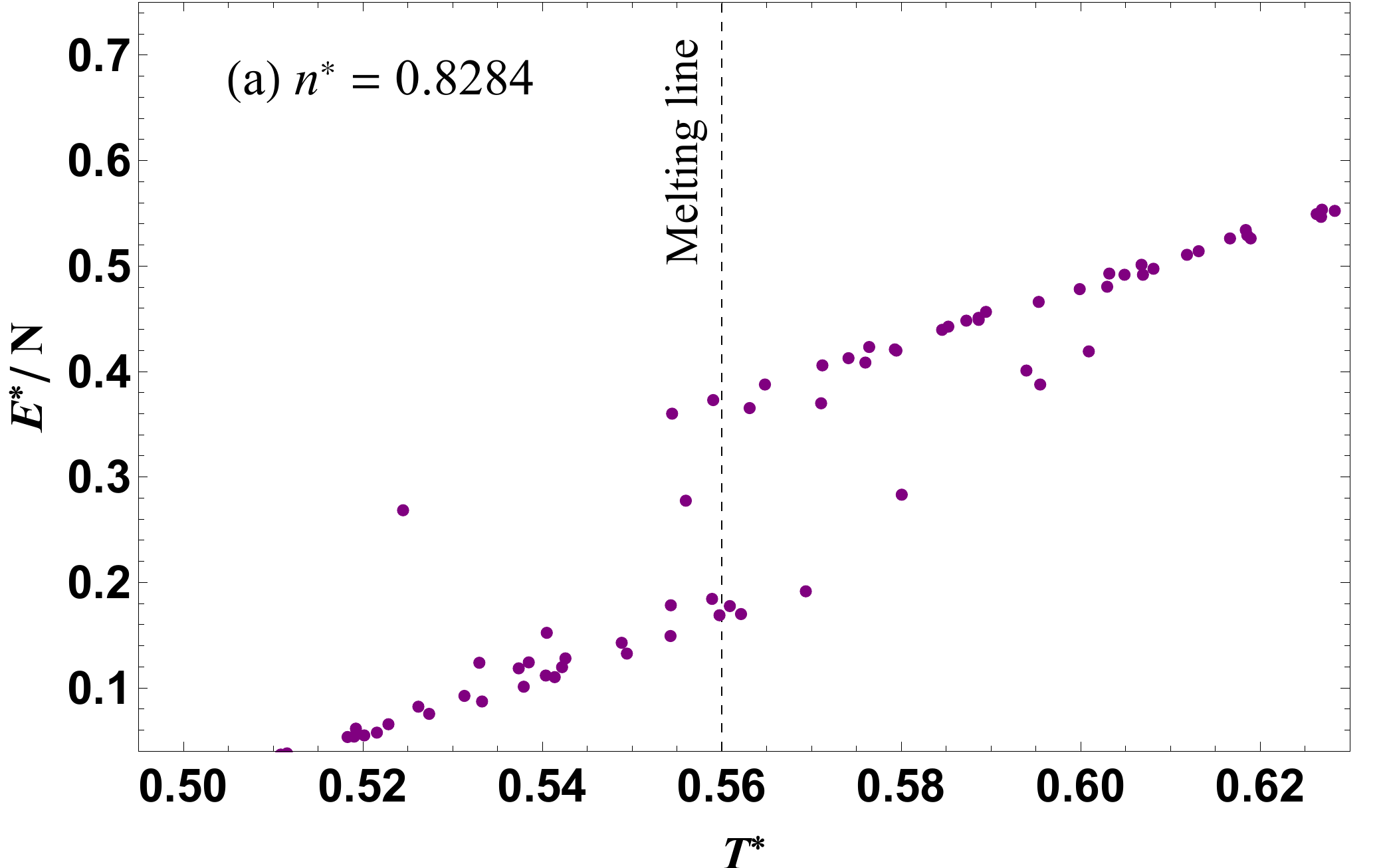}}}
{\scalebox{0.41}{\includegraphics{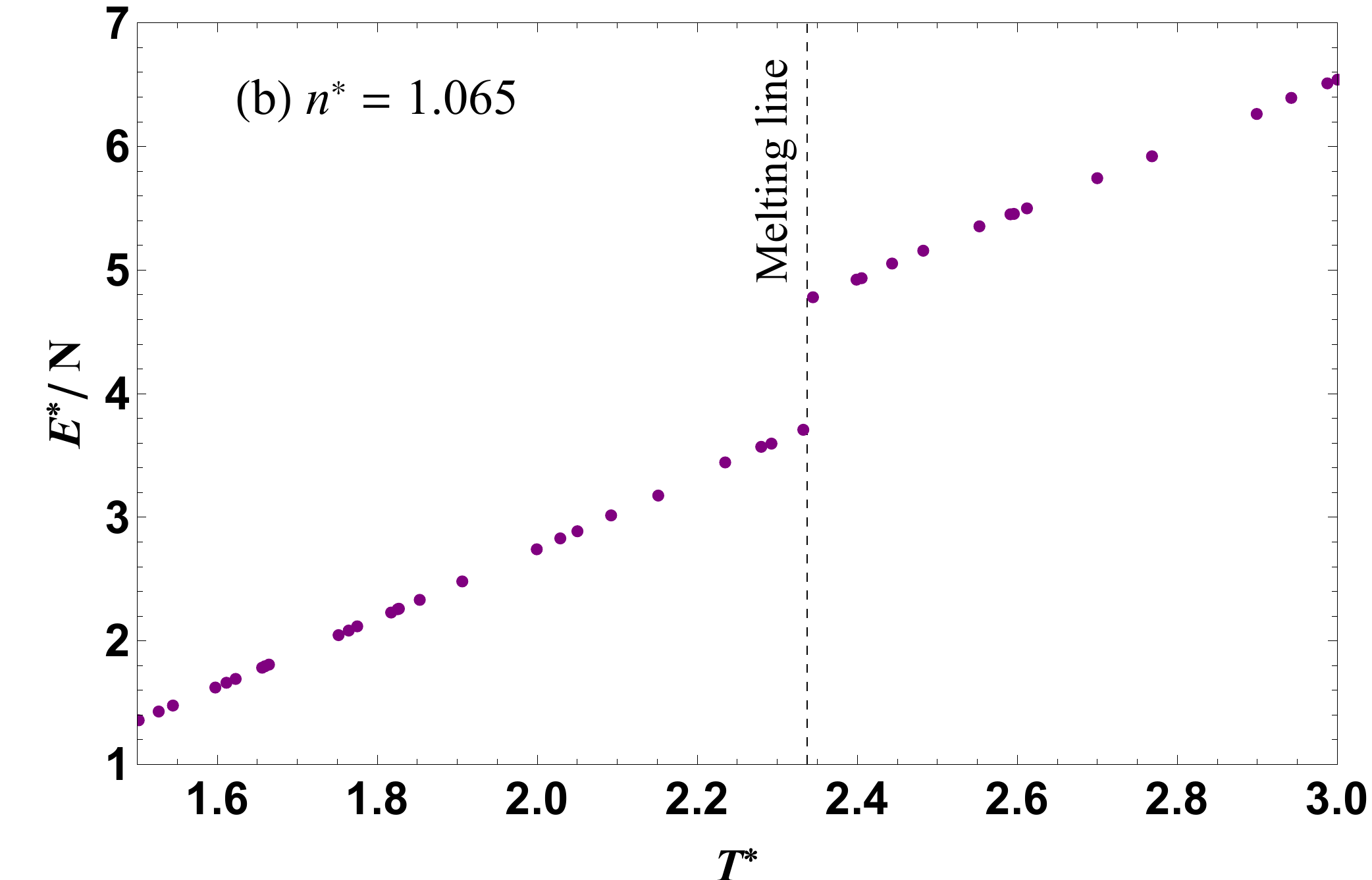}}}
\end{center}
\caption{Reduced energy per particle $E^*/N$ across the melting line at reduced concentrations of (a) 0.8284 and (b) 1.065.}
\label{fig:enptplot}
\end{figure}

\begin{figure}
\begin{center}
{\scalebox{0.41}{\includegraphics{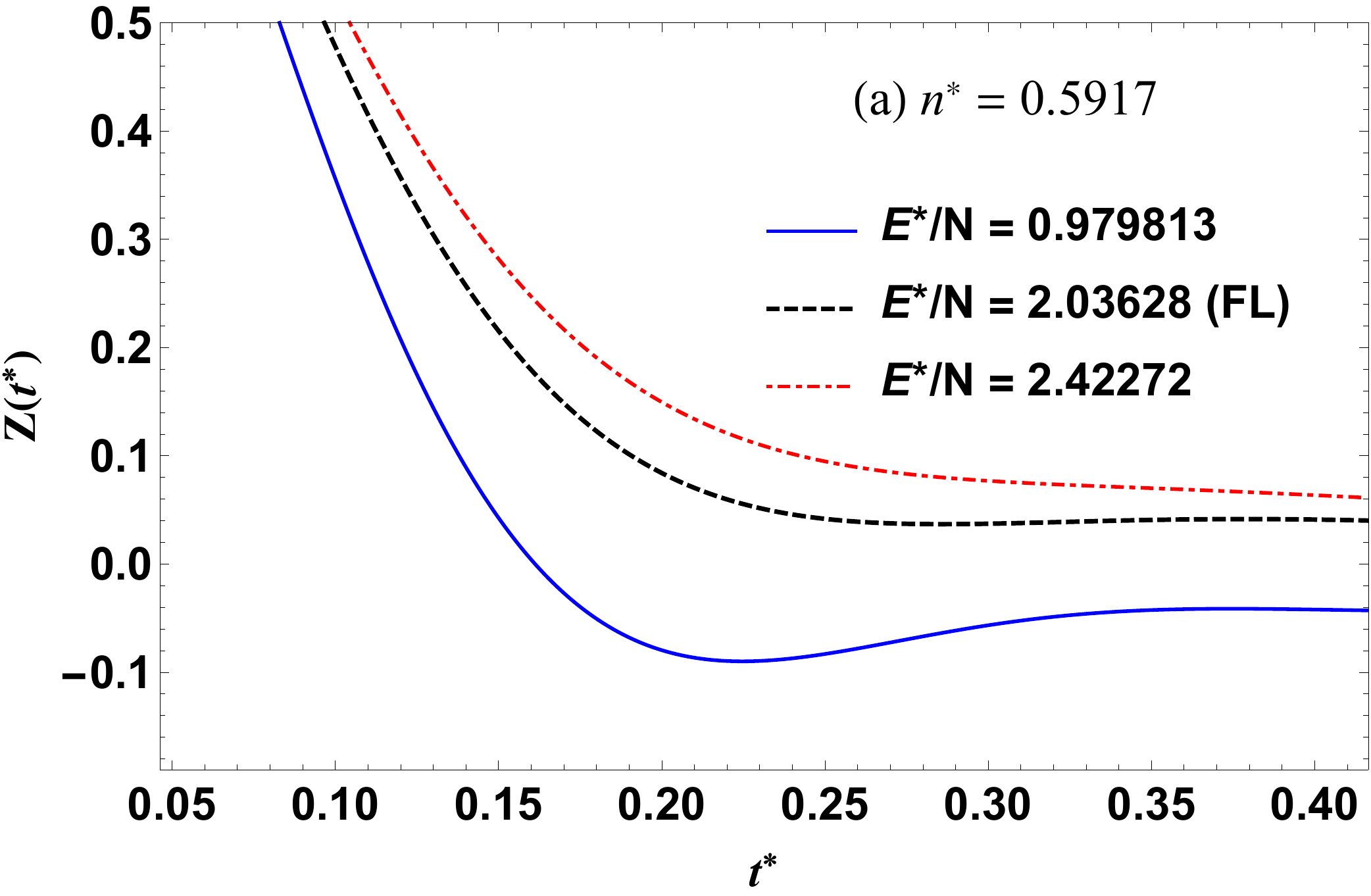}}}
{\scalebox{0.41}{\includegraphics{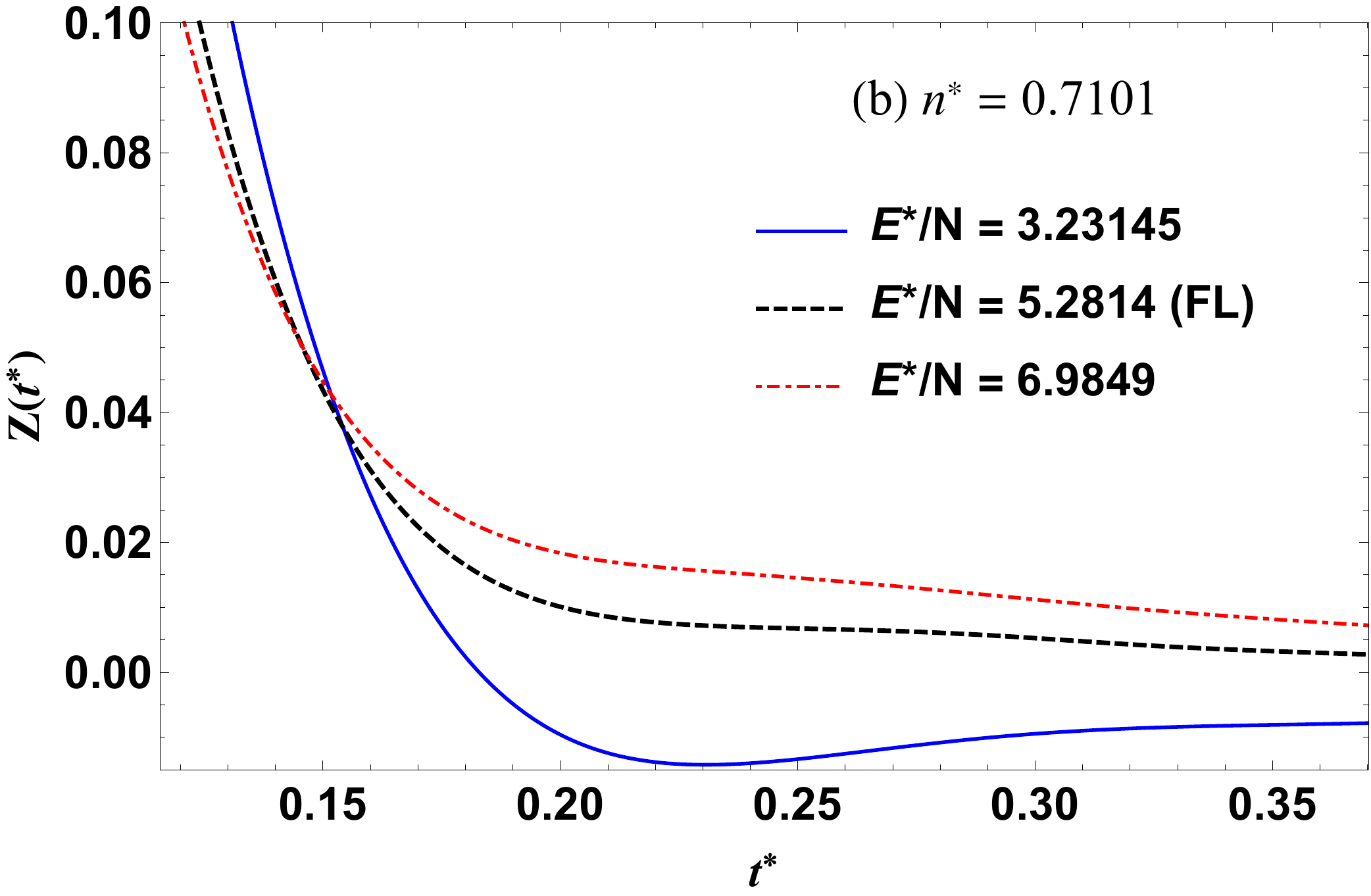}}}
{\scalebox{0.41}{\includegraphics{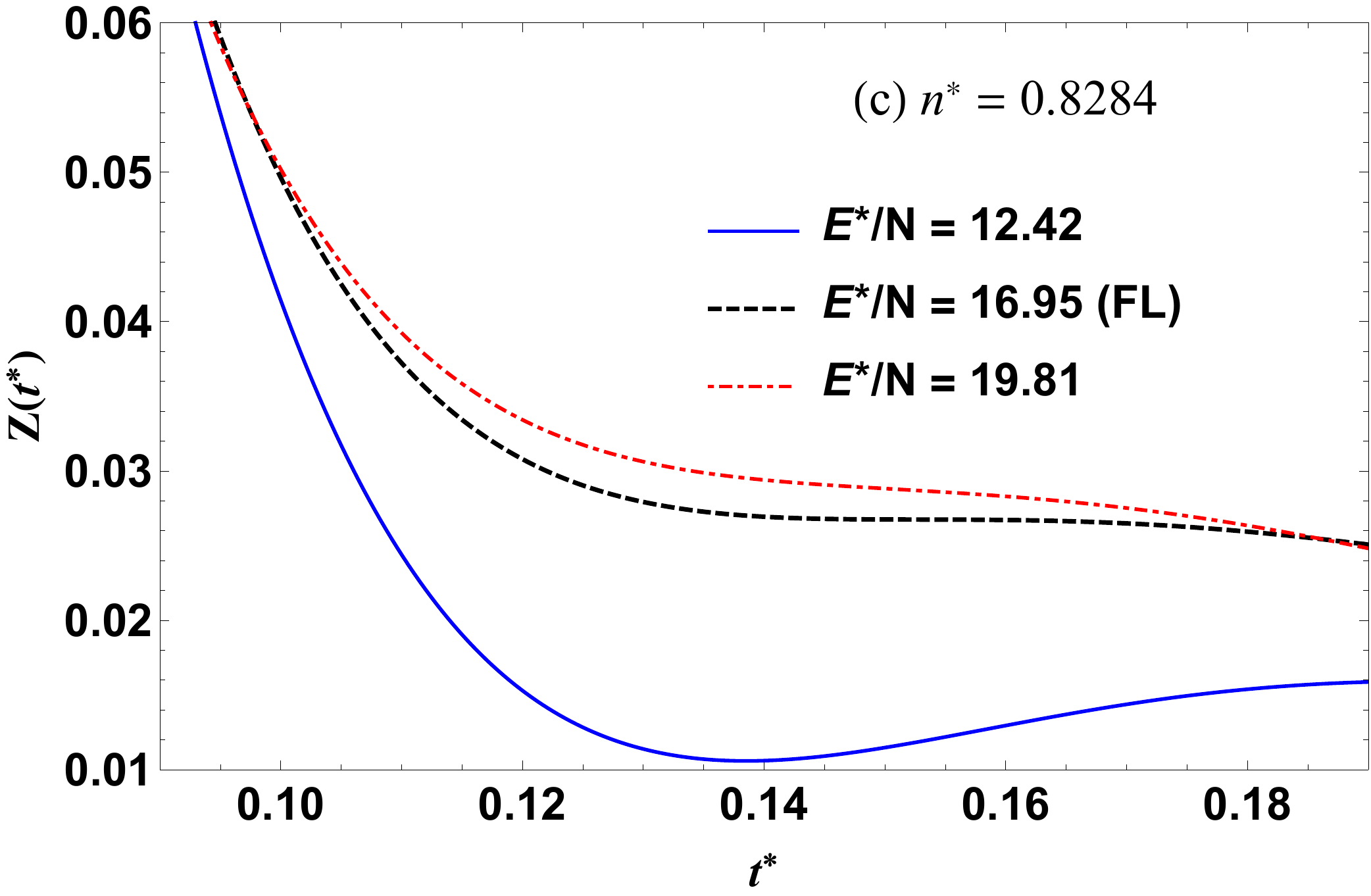}}}
{\scalebox{0.41}{\includegraphics{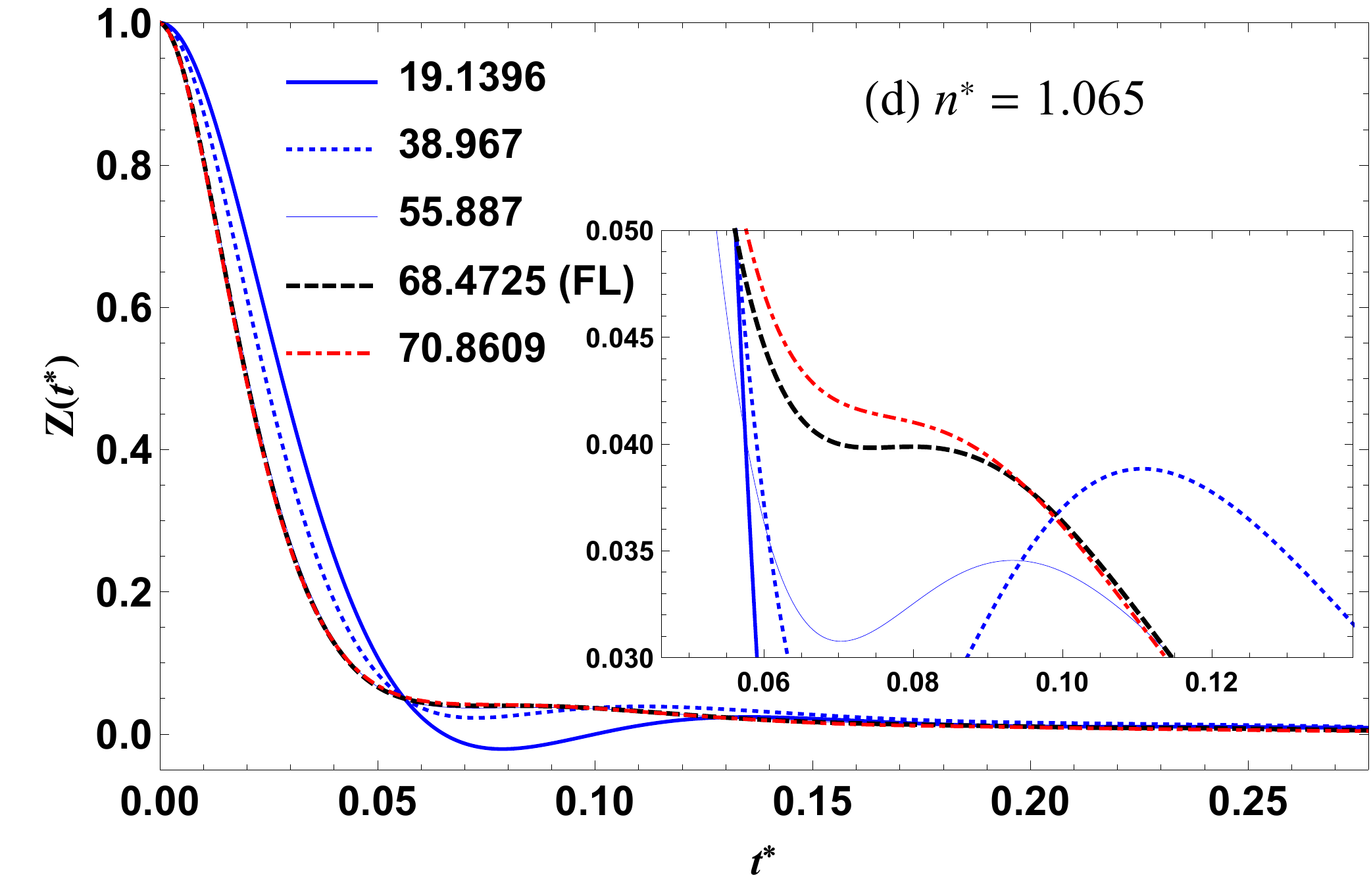}}}
\end{center}
\caption{Velocity autocorrelation functions as a function of reduced energy per atom and reduced time $Z(t^*)$ at the four different concentrations.}
\label{fig:vafs}
\end{figure}

\begin{table}
\begin{center}
\begin{tabular}{ |c|c|c|c|c| } 
\hline
$n^*$ & 0.5917 & 0.7101 & 0.8284 & 1.065 \\
 \hline
    $\rho$ (g/ml) & 0.9991 & 1.199 & 1.403 & 1.798 \\
    $n$ (\AA$^{-3}$) & 0.01506 & 0.01807 & 0.02108 & 0.02710 \\   
    $T_{\rm{F}}$ (K) & 82 & 295 & 997 & 3850 \\
    $T^{*}_{\rm{F}}$ & 0.70 & 2.45 & 8.33 & 32 \\
    $E_{\rm{F}}/N$ (eV) & -0.08 & 0.0 & 0.116 & 0.650\\
    $E^{*}_{\rm{F}}/N$ & 2.05 & 5.28 & 16.9 & 69.0\\
 \hline
\end{tabular}
\caption{Thermodynamic data for each system investigated: $\rho$ - mass density; $n$ concentration (number density); $T_{\rm{F}}$ - temperature at the Frenkel line; $T^{*}_{\rm{F}} = k_{\rm{B}} T_{\rm{F}} / \epsilon$ - reduced temperature at the Frenkel line;  $E_{\rm{F}}/N$ - energy per particle at the Frenkel line; $E^{*}_{\rm{F}}/N = E_{\rm{F}}/ \epsilon N$ - reduced energy per particle at the Frenkel line. $k_{\rm{B}}$ is Boltzmann's constant and $\epsilon$ is given in Tab. \ref{tab:ljparams}. The reference energy is $E^*/N = 0$ at $n^{*} = 0.5917$ and $T^{*} = 0.5$ (20 K).}
\label{tab:tdparams}
\end{center}
\end{table}

We plot VAFs in Fig. \ref{fig:vafs}, indicating the FL determined by the disappearance of the minima. At lower densities, the disappearance of the minima happens fairly steadily. However, at the highest density, a very slightly minimum remains for a temperature range that spans almost 1000 K. The ``zoomed-in" inset of Fig. \ref{fig:vafs} shows the gradual disappearance of this minimum - these VAFs are mostly indistinguishable at a lower resulotion despite at very different energies. This means that after most atomic oscillation is dispersed, a very slight component remains disappears far more gradually, which happens because the system remains fixed at a high density. In this sense, the system is almost completely diffusive and ``gas-like" far before the disappearance of the minimum, and the last leg of the transformation takes place much more slowly. Energies and temperatures at the FL are listed in Tab. \ref{tab:tdparams}.

\begin{figure}
\begin{center}
{\scalebox{0.41}{\includegraphics{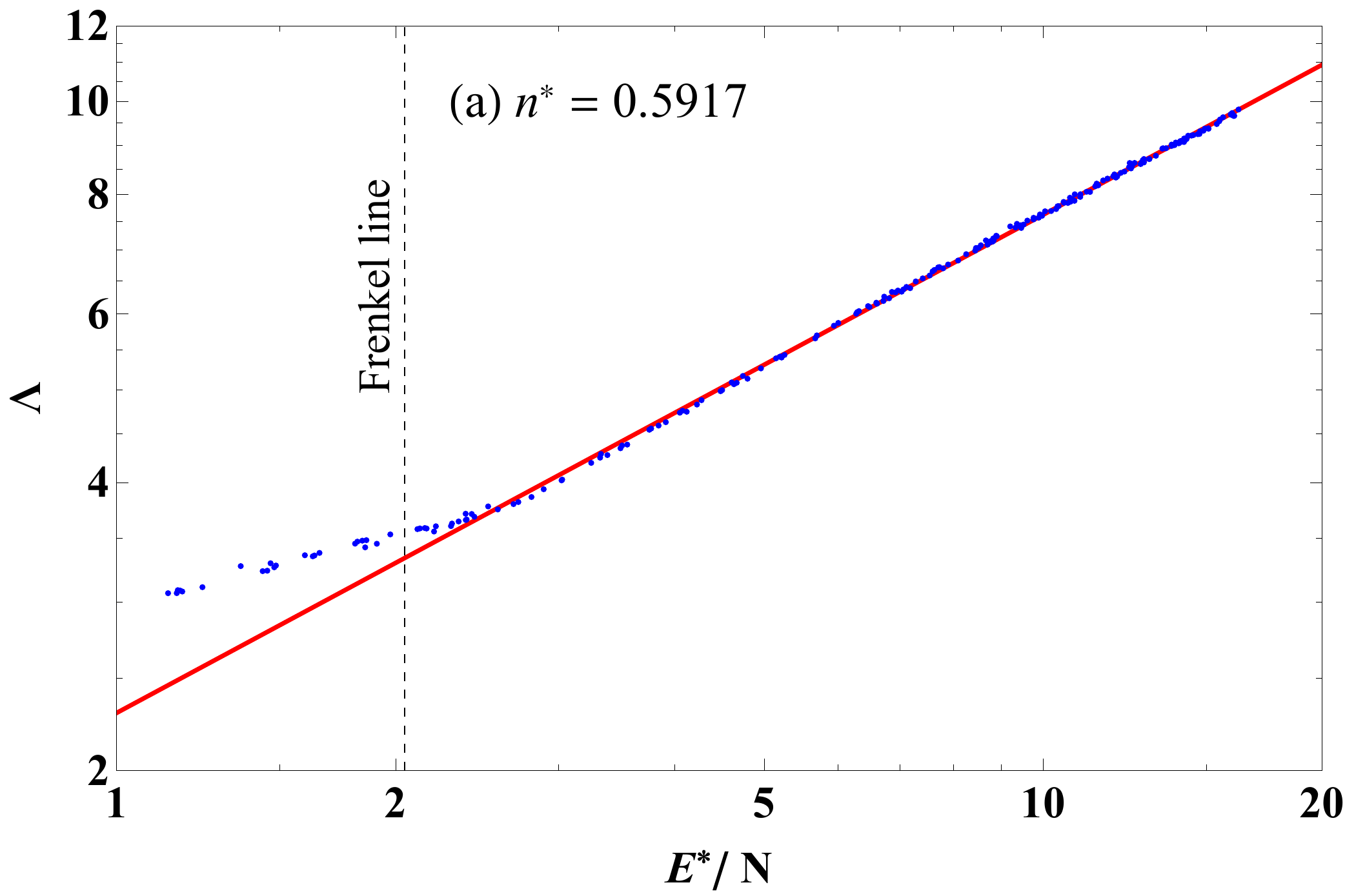}}}
{\scalebox{0.41}{\includegraphics{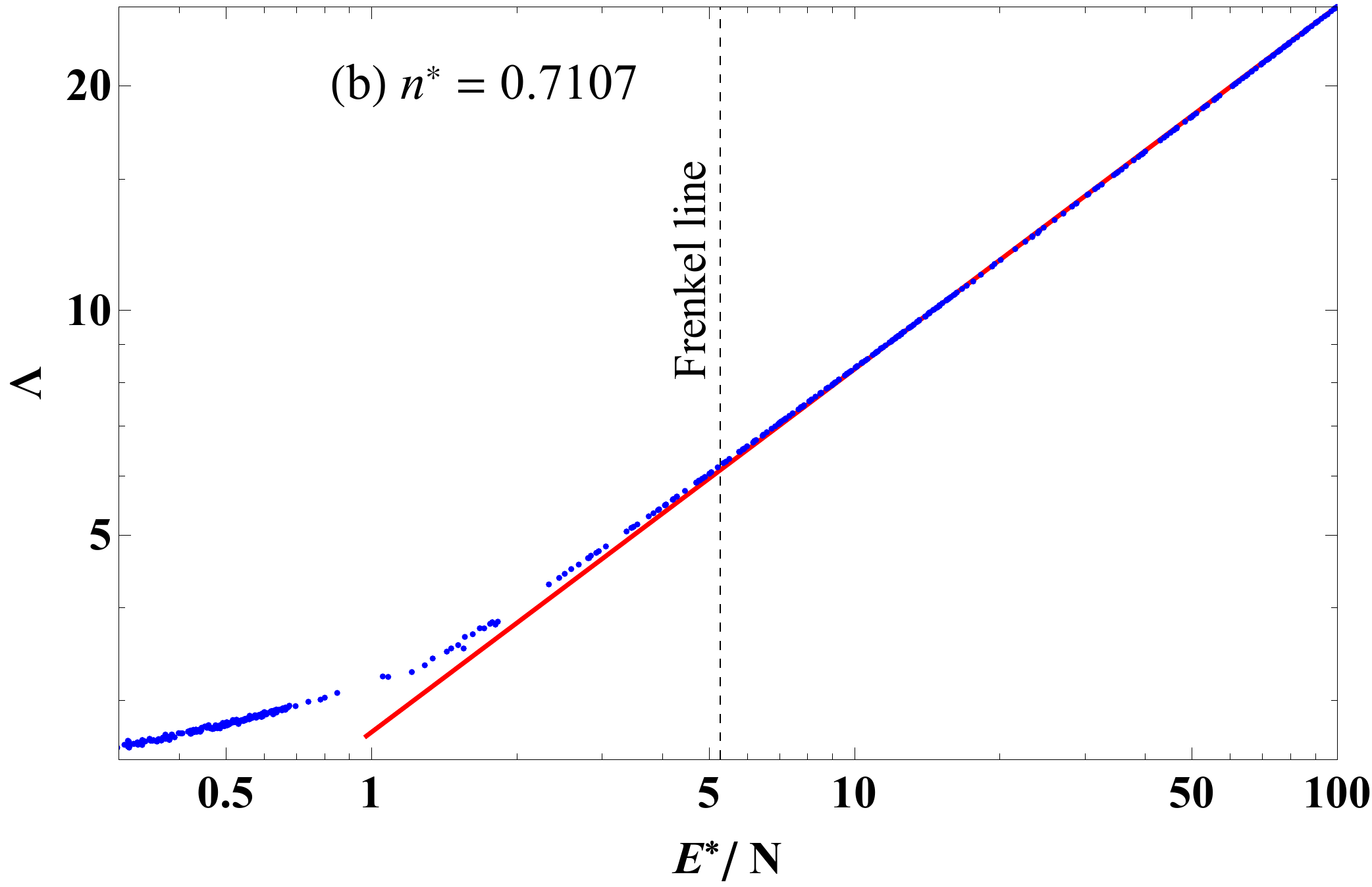}}}
{\scalebox{0.41}{\includegraphics{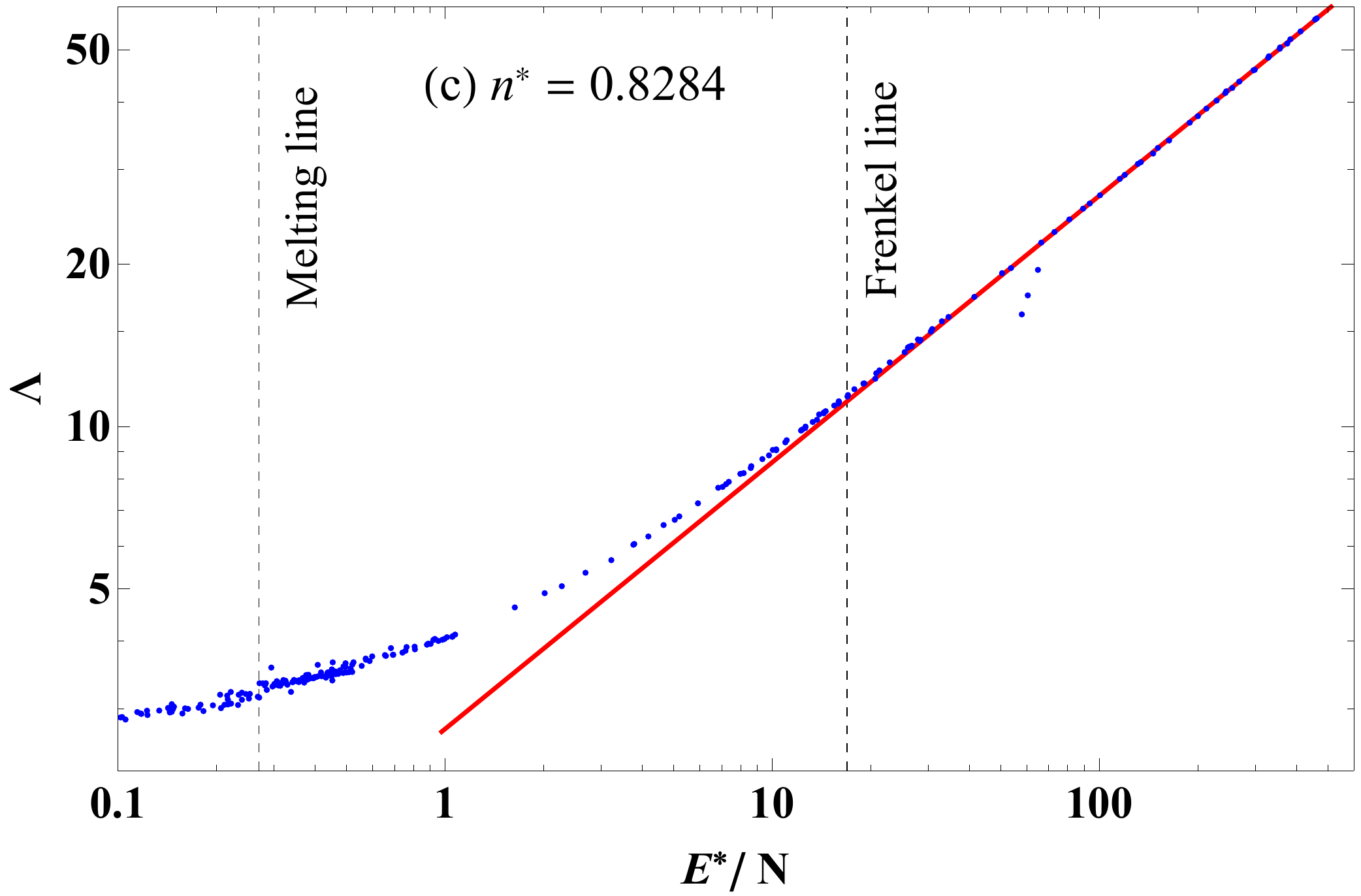}}}
{\scalebox{0.41}{\includegraphics{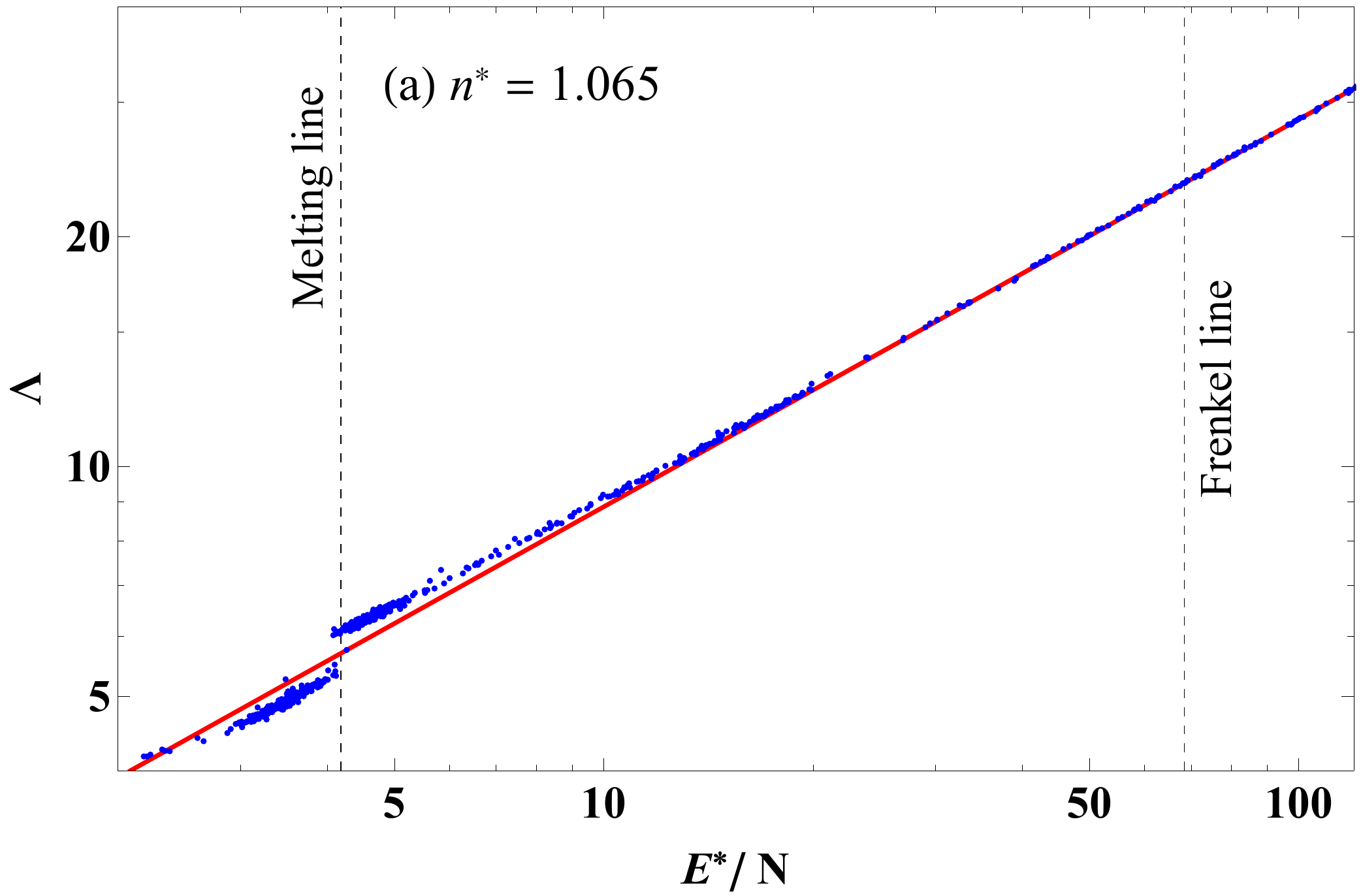}}}
\end{center}
\caption{Maximal Lyapunov exponent $\Lambda$ in the fluid state at reduced concentrations of (a) 0.5917; (b) 0.7107; (c) 0.8284; (d) 1.056. The red lines are fitted power-law relationships $\Lambda = a (E^*)^{b}$, meant as visual guides.}
\label{fig:lmefl}
\end{figure}

Fig. \ref{fig:lmefl} plots the MLE as a function of reduced energy, up to and beyond the FL. The high-temperature functional dependence of the MLE is clearly visible with the logarithmic axes: $\Lambda = a (E^*)^b$. At lower densities, the crossover to this power-law relationship closely coincides with the dynamical crossover at the FL. At the highest density, the dynamical crossover occurs deep within this power-law regime. However we note, as discussed above, a very small minimum in the VAFs disappears in the energy range of $E^*/N \approx 40$ to $E^*/N \approx 68$ (this is a larger range than that between the melting line and FL at the other densities), which corresponds to a very minor component of molecular oscillation disappearing in this range. For the most part, atomic oscillation gives way to diffusion at much lower energies than the disappearance of the minimum, represented in Fig. \ref{fig:lmefl} by the gradual approach to the power-law relationship as oscillatory modes disappear. 

The reason the crossover in the MLE at the FL is gradual rather than abrupt is because the crossover in dynmamics is also gradual. Across the melting line, the particle dynamics abruptly gain an oscillatory character. A liquid just above the melting line has a finite relaxation time between diffusive ``jumps", allowing it to support transverse collective modes below a certain wavelength \cite{Boon1991, Balucani1994, Hansen2006}. As temperature is increased, the relaxation time becomes shorter, reducing the maximum wavelength of transverse modes and thereby decreasing the heat capacity due to a reduction of the degrees of freedom in the system \cite{Bolmatov2012}. As the relaxation time drops below the oscillation period (at the FL), all oscillatory motion is lost, the system becomes fully diffusive, and the transverse spectrum becomes empty. This is accompanied by a thermodynamic crossover. Below the FL, the decrease in heat capacity is caused by the loss of long wavelength transverse modes due to the increasing relaxation time, above the FL, the decrease in heat capacity is caused by the loss of short wavelength longitudinal modes as the mean free path increases \cite{Trachenko2016}. In harmonic systems, this crossover takes place at a heat capacity of $c_{\mathrm{V}} = 2 k_{\mathrm{B}}$. The thermodynamic and dynamic (VAF) criteria give the same line on the phase diagram \cite{Brazhkin2013, Kryuchkov}. 

This interpretation allows us to make sense of our results here. Between the melting line and the FL, as the relaxation time decreases, the MLE increases with energy. The MLE increases because of the increased prevalence of diffusion events. Diffusion events involve an abrupt change in phase space coordinates as an atom escapes from its local ``cage" into another (see Fig. \ref{fig:traj}). These events are typically instigated by an atom's neighbours opening a low-energy pathway to form a neighbouring cage with their thermal motion. We propose that these diffusion events are the liquid equivalent to ``collisions" from kinetic theory because they involve a near-instantaneous decorrelation in particle coordinates and velocities and are very sensitive to initial conditions. The discontinuity of the MLE at the melting line is due to the sudden introduction of these events. This regime terminates smoothly as the relaxation time becomes comparable to the liquid oscillation period and a local rigid structure can no longer be defined. In other words the state becomes dynamically pure as atoms are continuously diffusing rather than doing so in opportunistic jumps (again see Fig. \ref{fig:traj}). The events of dynamic sensitivity are now the collisions of kinetic theory. Scattering is what makes the Lorentz gas a chaotic dynamical system \cite{Bunimovich1980}. These collisions now determine the evolution of the MLE without contribution from diffusion events, which is why it follows a single functional form. The collisions become more frequent with temperature at a fixed density. For a hard-sphere gas, the mean collision rate is \cite{Blundell2010}
\
\begin{equation}
    \label{eq:collisiontime}
    w_{\mathrm{coll}} = n \pi d^2 \sqrt{\frac{6 k_{\rm{B}} T}{m}}.
\end{equation}
\
This is a concave function of temperature (and thus energy), which is a property exhibited by the MLE at all densities (the gradient in the log-log plots in Fig. \ref{fig:lmefl} is less than unity). At the lower densities, this power-law regime spans more than an order of magnitude of energy above the FL. The fluid at the highest density, even well below the FL, is mostly dominated by diffusion and collisions, but there is a transitory period of oscillation for some molecules. We can interpret that collisions are responsible for the bulk of dynamical instability in these states, but a small fraction of atoms at any given time undergoing oscillation do not contribute to dynamical instability in this way. This crossover period of small deviation from the power law is much smaller at lower densities.

\section{Conclusions}

We have presented a novel study of Lyapunov exponents, focusing on the supercritical fluid state. We find that the MLE in the ``gas-like" deeply supercritical LJ system evolves with energy according to a single analytic function, which we explain in terms the fluid's dynamical evolution. Recent advances in the field of theoretical liquid physics \cite{Trachenko2016} have explained many liquid properties by describing the state in terms of two dynamical modes: molecular oscillation around quasi-equilibrium positions, and abrupt diffusion events between quasi-equilibrium positions. Molecular oscillation terminates at the FL, and the dynamical evolution switches from a loss of oscillation to a decline of collisions. This dynamical crossover causes a crossover in both thermodynamics and structure in many different fluid systems \cite{Wang2017,Wang2019,Prescher2017,Smith2017,Proctor2019,Cockrell2020,Cockrell2020a}. On the basis of our MD simulations, this same dynamical crossover causes a crossover in the MLE. We explain this crossover in terms of diffusion events and collision events, prevalent below and above the FL respectively, which we propose are the major contributors to dynamical stability in these fluid states. The Lyapunov spectrum is linked to dynamics much more intimately than thermodynamics and structure, and has been used in the past to indicate changes of phase \cite{Nayak1995, Mehra1997, Dellago1997,Kwon1997,Butera1987,Nayak1998,Barre2001}. Our results therefore do not only help understand microscopic chaos in the fluid state, but also show that the depiction of liquids as dynamically mixed states and the idea of the FL are supported directly by properties of the classical phase space itself.
\bibliography{collection}{}
\bibliographystyle{unsrt}
\end{document}